\documentclass[showpacs,twocolumn,amsmath,amssymb,floatfix,aps,prl]{revtex4}

\usepackage{graphicx}
\usepackage{dcolumn}
\usepackage{bm}

\begin{document}

\title{Observation of Cs Rydberg atom macrodimers}

\author{K. R. Overstreet}
\author{A. Schwettmann}
\author{J. Tallant}
\author{D. Booth}
\author{J. P. Shaffer}
\affiliation{University of Oklahoma, Homer L. Dodge Department of
Physics and Astronomy, 440 W. Brooks St., Norman, OK 73019}

\date{\today}

\begin{abstract}
We report the observation of cold Cs Rydberg atom molecules bound at
internuclear separations of $R\sim$3-9$\,\mu$m. The bound states
result from avoided crossings between Rydberg atom pair interaction
potentials in an applied electric field. The molecular states can be
modified by changing the applied electric field. The molecules are
observed by mapping the radial separation of the two Rydberg atoms
as a function of time delay between excitation and detection using
the Coulomb repulsion of the ions after pulsed field ionization.
Measurements were performed for $63D+65D$, $64D+66D$, $65D+67D$, and
$66D+68D$ pairs. The experiment is in good agreement with
calculations of the pair interactions for these states.
\end{abstract}


\maketitle

\begin{figure}
\includegraphics[width=3.375in]{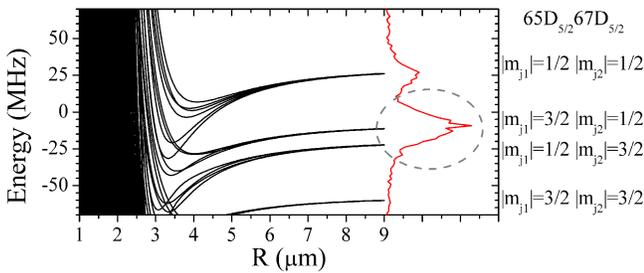}
\caption{\label{fig:1} Integrated atomic ion yield spectra and pair
potentials for $65D+67D$ with $\epsilon \,= \,190\,$mV$\,$cm$^{-1}$.
The excitation laser intensity is $\sim 500\,$W$\,$cm$^{-2}$. All
fine structure and $\Omega$ are plotted. $\Omega
\,$=$\,$m$_{j1}$+m$_{j2}$ is the projection of the angular momentum
on $R$. The dashed line highlights one of the features studied in
this paper.}
\end{figure}

Frozen Rydberg gases \cite{Mourachko97,Anderson97} have been the
subject of intense research recently.  The construction of fast
quantum gates and single photon sources using dipole blockade
\cite{Gould04,Weidemuller04,Raithel05,Pillet06,Pfau07,Jaksch00,Lukin01,Saffmann08},
the study of cold Rydberg atom molecules
\cite{Green00,Boisseau02,Schwettmann07,Bendkowsky09}, and the
investigation of many body physics \cite{Mourachko97,Anderson97} are
central motivations for this work. Cold Rydberg atom molecules are
exciting because of the interesting properties that these objects
possess. Molecules formed by two cold Rydberg atoms are called
macrodimers since the atoms are bound at distances $> 1 \,\mu$m. It
has been suggested that due to their delicate nature, macrodimers
can be used to study vacuum fluctuations, quenching in ultracold
collisions \cite{Boisseau02} and Rydberg atom interactions including
their controllability with applied electric fields. Prior
experiments have been unable to unambiguously confirm that these
unique states of matter exist as bound states
\cite{Farooqi03,Overstreet07}. We report the first experimental
observation of bound macrodimers which have one of the largest, if
not the largest, molecular bond observed to date.

The macrodimers that we observe result from avoided crossings
between Rydberg atom pair interaction potentials in an applied
electric field, $\epsilon$ \cite{Schwettmann07}. These macrodimers
are unique because they feature a quasi-continuum of bound states,
are found at extremely large internuclear separation,
$R\sim$3-9$\,\mu$m, and are formed by the interplay between Van der
Waals interactions and the Stark effect resulting from $\epsilon$.
$\epsilon$ can be used to stabilize, destroy or modify the potential
well that gives rise to the bound states. A spectrum and calculated
potentials \cite{Schwettmann06} for the $65D+67D$ pair with
$\epsilon = 190\,$mV$\,$cm$^{-1}$, are shown in Fig. \ref{fig:1}.
The potentials have prominent wells at $R\sim$3-9$\,\mu$m.  These
potential wells support hundreds of bound states
\cite{Schwettmann07} with maximum energy spacings of
$\sim$100$\,$kHz. The lifetimes of the molecules are limited by the
radiative and blackbody decay of the atoms \cite{Schwettmann07}.

\section{Detecting Macrodimers}

Macrodimers are difficult to detect. Pulsed field ionization (PFI)
breaks the molecule apart and complicates time-of-flight (TOF) mass
spectrometry. Using selective field ionization to ionize only one of
the Rydberg atoms forming the molecule creates an ion close to the
remaining Rydberg atom. The electric field of the ion can ionize the
atom. Field ionizing both atoms creates two ions that repel each
other, causing the complex to dissociate. All these possibilities
lead to the detection of singly ionized Cs atoms at the same average
TOF. Detection of the molecular ro-vibrational spectrum is difficult
because it is a quasi-continuum. For the state shown in Fig.
\ref{fig:1}, the rotational constant is $\sim 6\,$Hz while the
maximum classical angular momentum at the equilibrium separation,
$R_e$, is $l_{max} = \mu v R_e/ \hbar \approx 310$. $\mu$ is the
reduced mass and $v$ is the thermal velocity. Given that the
radiative decay time leads to a natural line width of $\sim 10\,$kHz
for 66$D$ and taking into account the allowed rotational transitions
due to the large impact parameter, each vibrational line will have a
width of $\sim 580\, $kHz. This width is much greater than the
vibrational energy spacings. Consequently, it is unlikely that the
ro-vibrational states of a macrodimer can be resolved unless deeper
and narrower potential wells can be located.

We use the recoil velocity distribution to detect the Cs macrodimers
in our experiment \cite{Overstreet07,Tallant06}. A basic
characteristic of a bound state is that the atoms remain in close
proximity for a long time compared to the vibrational period. The
signature of molecule formation in our experiment is based on this
principle. We measure the recoil velocity of the ion fragments after
PFI as a function of the delay, $\tau$, between excitation and PFI.
If two Rydberg atoms are close together and pulsed field ionized,
the ions created will Coulomb repel each other. The amount of
Coulomb repulsion, which is reflected in the recoil velocity of the
ions, is a measure of the $R$ at which the pair was ionized.
Measuring the recoil velocity of the ions as a function of $\tau$
maps the distribution of radial separations of the two Rydberg atoms
as a function of time. For a bound state, the distribution of $R$
will remain relatively constant as a function of $\tau$. For a
dissociative process, the distribution of $R$ will change with
$\tau$. The recoil velocity can be measured by recording the TOF
distribution of the ions that were created by PFI of the molecule at
a detector in a TOF experiment \cite{Tallant06}.

In our experiment, we do not expect to observe oscillations of the
Coulomb repulsion due to molecular ro-vibrational wave-packet
motion. The vibrational periods are $\lesssim 10\,\mu$s.
Experimental limitations such as long term laser stability and
excitation pulse average away any oscillations that may be present.
Future experiments designed to observe wave-packet motion will be
interesting since macrodimers exist at the interface between quantum
and classical dynamics.

\section{Experimental Methods and Results}

\begin{figure}
\includegraphics[width=3.375in]{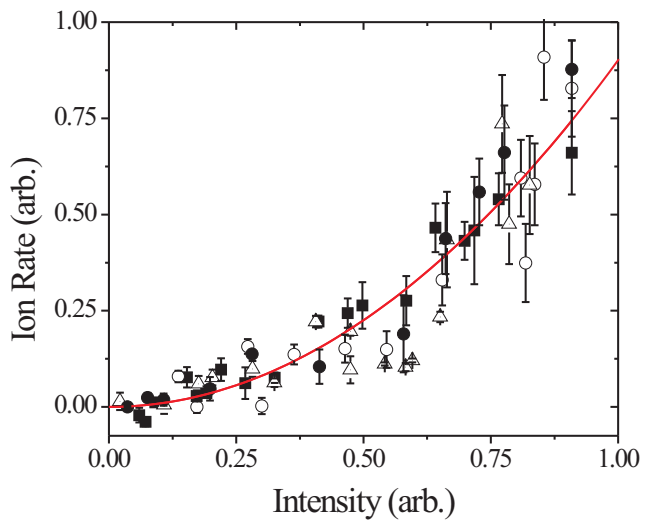}
\caption{\label{fig:2} The rate dependence on excitation laser
intensity.  Solid triangles ($\blacktriangle$) denote 63D+65D
molecular state, open triangles ($\vartriangle$) 64D+66D, open
circles ($\circ$) the 65D+67D, and closed squares ($\blacksquare$)
66D+68D. Intensities and rates have been normalized to show the
quadratic dependence. The solid line is a quadratic fit to the
data.}
\end{figure}

The macrodimers are created in a Cs magneto-optic trap (MOT). The
MOT contains $10^{7}\,$ atoms at a density of $\rho\,\sim\,2
\times10^{10}\,$cm$^{-3}$ at a temperature of $\sim124\,\mu$K
\cite{Tallant06}.  The background pressure is $\sim \,5
\times10^{-10}\,$Torr. The MOT is formed between two electric field
plates, $26.7\,$cm above a microchannel plate (MCP) detector. The
experimental apparatus is described in more detail in
\cite{Tallant06}.

The Cs macrodimers are excited by a 2-photon scheme
\begin{equation}{\label{eqn:1}}
6P_{3/2}+6P_{3/2}+2h\nu\rightarrow (n-1)D+(n+1)D,
\end{equation}
where $n\,$=$\,$64, 65, 66, and 67. The MOT trapping light is shut
off during the time when the molecules are excited and detected. The
MOT repumping light is on during the experiment. The atoms in the
$6P_{3/2}(F=5)$ state are excited from the $6S_{1/2}(F=4)$ state
using a 1$\,$mm spot size beam derived from the trapping laser. This
beam is crossed at 22.5$^\circ$ with a beam from a dye laser
(Coherent 699) operating at $\lambda=508-509\,$nm that drives the
2-photon transition in Eq. \ref{eqn:1}. The dye laser is focused to
a $\sim 25 \pm 10 \,\mu$m spot size that was measured with a CCD
camera. The dye laser polarization is parallel to the TOF axis. The
crossed beam geometry helps reduce background counts. The intensity
of the dye laser and the trapping light used to excite the atoms to
$6P_{3/2}(F=5)$ is 500$\,$W$\,$cm$^{-2}$ and 0.5$\,$W$\,$cm$^{-2}$
respectively.

Bound or free Rydberg atoms are detected in the experiment using
PFI. The experiment begins when the trapping light is switched off.
$2.5\,\mu$s after this light is switched off, the Rydberg excitation
beams are turned on for $5\,\mu$s. After a variable delay, $\tau$, a
PFI pulse is applied for $2\,\mu$s at an amplitude of
$76\,$V$\,$cm$^{-1}$ to ionize the Rydberg atoms that were excited
and push the ions to the MCP. The rise time of the electric field
pulse is $\sim 80\,$ns. A $153\,$mV$\,$cm$^{-1}$\,$\mu$s$^{-1}$,
$5\,\mu$s duration ramp electric field is applied immediately
preceding the PFI pulse to clear any stray ions from the excitation
region. The atoms move $< \,500\,$nm due to thermal motion during
this time. The ionized atoms have a TOF distribution centered at
$\sim 27\,\mu $s. The pulses from the MCP are processed by a
constant fraction discriminator and accumulated by a multichannel
analyzer. A constant voltage is applied to the upper field plate
during excitation to generate $\epsilon$. To measure TOF velocity
distributions, the sequence was repeated at a rate of $1\,$kHz which
resulted in count rates of $\sim \, 10\,$Hz.

Under the experimental conditions described, many molecular
resonances are observed for Rydberg atom pairs at different
$\epsilon$ \cite{Schwettmann07,Overstreet07}. In this experiment, we
measured molecular resonances for $63D+65D$, $64D+66D$, $65D+67D$
and $66D+68D$ pairs. We focused on these states because our
calculations indicated that bound states existed for certain
$\epsilon$ and these states satisfied the necessary experimental
requirements. The experimental considerations are a compromise
between an acceptable oscillator strength and the elimination of
background counts from Rydberg atoms. The oscillator strength for
the molecular excitation is predominantly due to mixing via
multipole interactions and electric fields with the nearest atomic
state which tends to decrease with increasing energy separation. The
asymptotic atomic states are far detuned from the excitation laser
frequency so without the mixing one expects little transition
strength to the molecular states. However, the noise in the
measurement increases with n as the molecular resonance moves closer
to the atomic line wing where some atomic Rydberg excitation occurs.

The excitation rate for each molecular resonance was first measured
as a function of excitation laser intensity and found to be
quadratic indicating a 2-photon excitation process, Fig.
\ref{fig:2}. These observations, our prior experiments
\cite{Overstreet07} and the fact that the gas is dilute are evidence
that the spectral features are from processes involving two Rydberg
atoms. The fact that the spectral features are associated with
molecular excitations does not necessarily indicate the presence of
macrodimers. Many resonances result from photo-initiated collisions
where the atoms are not excited to bound states \cite{Overstreet07}.

\begin{figure}[t]
\includegraphics[width=3.0in]{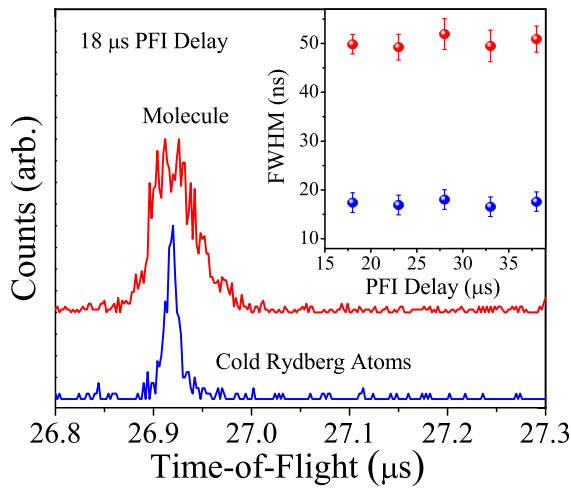}
\caption{\label{fig:3} Comparison of TOF distributions for the 66D
atomic case (blue) and the 65D+67D molecular case (red) at $\tau =
18 \,\mu$s. The horizontal axis is centered on the atomic TOF. These
distributions measure the particle velocities or kinetic energies
\cite{Tallant06}. The inset shows the full width at half maximum of
a Gaussian fit to each of the different distributions at different
time delays. The atomic data was taken by tuning the laser to the
atomic state under conditions of single particle counting.}
\end{figure}

Two sample TOF distributions at $\tau\,=\,18\,\mu$s are shown in
Fig. \ref{fig:3}. The inset of Fig. \ref{fig:3} shows the full width
at half maximum of a Gaussian fit to each of the distributions for
$\tau\,=\,18-38\,\mu$s. The narrow distribution was obtained by
tuning the excitation laser to the 66$D$ atomic resonance under
conditions for single atom counting. The broad distribution was
acquired by tuning the excitation laser to the feature circled in
Fig. \ref{fig:1}. The difference between the two distributions in
Fig. \ref{fig:3} is a result of the dynamics taking place when
single atoms versus pairs of atoms are excited. Rydberg atoms expand
from the excitation volume at the thermal velocity, $\sim8.9\,$cm
s$^{-1}$ for Cs. Cs atom pairs expand at the vector sum of their
thermal velocity, collisional recoil velocity and the velocity
acquired from Coulomb repulsion after PFI. The molecular signal is
much broader than the atomic signal, characteristic of the Coulomb
repulsion between the ions created at close range, $R \lesssim 10
\,\mu$m, by PFI. Although the amplitude of the traces shown in Fig.
\ref{fig:3} is similar, the count rates for the atomic signal are
3-4 orders of magnitude larger for identical excitation laser
intensities.

\begin{figure}[b]
\includegraphics[width=3.0in]{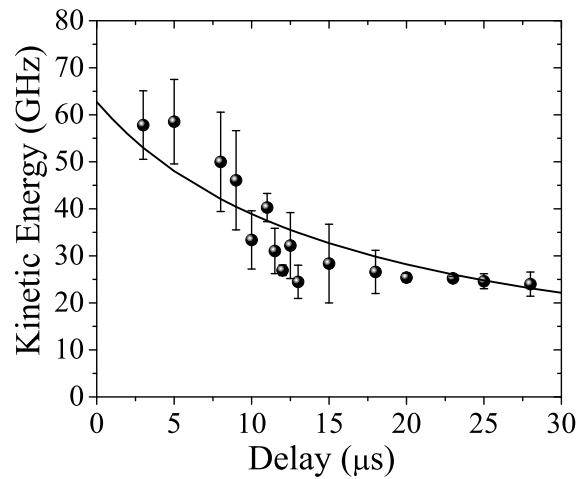}
\caption{\label{fig:4} This figure shows a plot of kinetic energy
vs. $\tau$ for a collision between two 89D atoms. The line is the
energy expected for a pair of Cs atoms starting at $R\,=\,2.8\,\mu$m
and moving uniformly apart at a velocity of 17$\,$cm$\,$s$^{-1}$ per
atom for a time $\tau$ before ionization as measured for this
collision in \cite{Overstreet07}.}
\end{figure}

The TOF distribution of a molecule will only be broadened by the
thermal velocity and the velocity obtained from Coulomb repulsion
because the collisional recoil velocity is zero. If the time between
PFI and the average arrival time at the detector is not changed, the
Coulomb repulsion between the atoms forming a molecule will remain
constant as a function of $\tau$ since the distribution of $R$
remains constant. The TOF distribution under these conditions
measures the distance between the excited atoms as a function of
$\tau$. This description is valid as long as there is no observable
wave-packet motion which would lead to an oscillating signal.

\begin{figure*}
\includegraphics[width=6in]{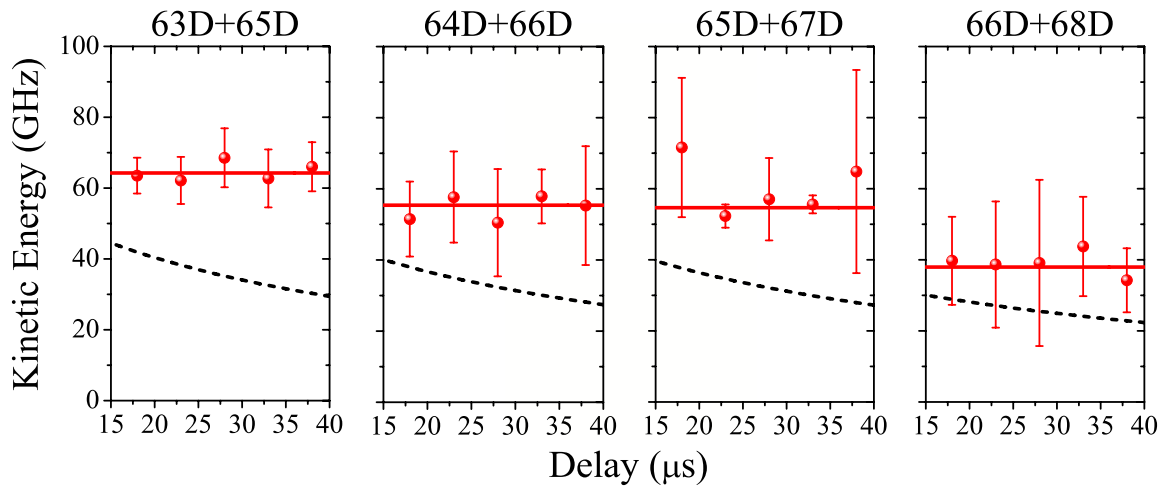}
\caption{\label{fig:5} Recoil kinetic energy vs. delay for $63D+65D$
with $\epsilon=224\,$mV$\,$cm$^{-1}$, $64D+66D$ with
$\epsilon=205\,$mV$\,$cm$^{-1}$, $65D+67D$ with
$\epsilon=190\,$mV$\,$cm$^{-1}$ and $66D+68D$ with
$\epsilon=158\,$mV$\,$cm$^{-1}$. Spheres denote the data taken for
the molecular resonances and the dotted black line shows a
calculation of the expected results of a binary collision with
thermal velocity as described in the text.}
\end{figure*}

Pairs of atoms undergoing a collision have increasing internuclear
separation, $R$, with increasing $\tau$ so the Coulomb repulsion
will decrease as $\tau$ increases for fixed TOF. An example of this
behavior is shown in Fig. \ref{fig:4} for the collision between 89D
atoms studied in \cite{Overstreet07}. These dynamics will result in
a Coulomb repulsion broadened TOF distribution at short $\tau$ that
becomes smaller as $R$ increases due to the exit channel velocity of
the fragments. Asymptotically, $R \gtrsim 15 \,\mu$m, when two atoms
leaving a collision are ionized, the Coulomb repulsion is negligible
and the velocity is determined by the exit channel velocity and
sample temperature alone \cite{Overstreet07}. Collisions where the
atoms are excited at long range and are allowed to evolve on
attractive molecular potentials can also be studied in this way but
this is a subject of future work.

To observe macrodimers, TOF distributions were acquired for
$\tau=18-38\,\mu$s.  Figure \ref{fig:5} shows the data for the 4
states used in the experiments. The times were chosen to be short
enough to avoid decay of one of the molecular partners, but long
enough to measure the dependence on $\tau$ so that a collision could
be distinguished from a macrodimer. The lifetime of a $\sim60D$
state including blackbody decay is $\sim100\,\mu$s
\cite{Farley81,He90}. Qualitatively, the kinetic energy distribution
is not changing with $\tau$ indicating that the spectral feature
that the laser is tuned to is a macrodimer.

A linear fit was applied to the data for each molecular pair shown
in Fig. \ref{fig:5} to determine the minimum collision velocity that
would give a similar TOF distribution. The velocity allowed by the
experimental error in the measured distributions is $0\pm1\,$cm
s$^{-1}$ for $63D+65D$, $-2.7\pm2.1\,$cm s$^{-1}$ for $64D+66D$,
$2.2\pm2.7\,$cm s$^{-1}$ for $65D+67D$ and $3.4\pm4\,$cm s$^{-1}$
for $66D+68D$. All errors quoted are standard deviations. The
collision velocities are all zero within the resolution of our
spectrometer, $\sim 2.5\,$cm$\,$s$^{-1}$, for all pairs
\cite{Overstreet07,Tallant06}. The noise in the measurement
increases with $n$ because the molecular states are getting
energetically closer to the nearest atomic line.

The data is also consistent with our calculations of the interatomic
potential wells. Notice that the Coulomb repulsion decreases with
increasing n. This is because the potential wells shift to larger R
as $n$ increases. The fact that the $64D+66D$ and $65D+67D$
macrodimers have approximately the same recoil kinetic energy is
also predicted by our calculations. The $R_e$ of the wells closest
to the spectral lines studied at the $\epsilon$ quoted in Fig.
\ref{fig:5} for $63D+65D$, $64D+66D$, $65D+67D$, and $66D+68D$ are
approximately 3.3$\,\mu$m, 3.5$\,\mu$m, 3.6$\,\mu$m and 3.9$\,\mu$m,
respectively. The $R_e$ are a measure of the potential well shape.
It should be pointed out, however, that most of the excitation
probability is at large $R$. This is due to the Frank-Condon factors
and the number of available pairs at particular $R$ at our
densities. These factors favor near dissociation states. The
macrodimers spend most of their time at $R\,=\,4-9\,\mu$m at the
outer turning points of the potential wells. We also note that there
is not a simple $n$ scaling law dependence to the well shape because
this depends on $\epsilon$ as well as n.

We compared the measured TOF distributions to simulations of the ion
trajectories in our apparatus. Each simulation includes the
spectrometer geometry, $\epsilon$, and the PFI pulse. $R$ for each
macrodimer was chosen uniformly from a range based on the calculated
potential wells. Molecules or atoms were placed at random in the
excitation volume according to the excitation laser intensity
distribution. The simulated TOF distribution width was in good
agreement with the molecular signal. The atomic measurement yields a
$\sim50\,\%$ larger TOF distribution most likely due to the
alignment of the dye laser focal spot to the MOT and the uncertainty
in determining the laser spot size. The Coulomb repulsion for our
range of $\tau$ is at least 3 times larger than the thermal
broadening so the spot size uncertainty does not prevent the
measurement. The expected behavior shown in Fig. \ref{fig:4} as a
solid line is also in good agreement with the data further verifying
the accuracy of the simulations.

We compared the experimental data on the macrodimers to the expected
behavior of an ultra-low energy, 5.22$\,$MHz, binary collision. The
recoil velocity of the atom pair was taken to be equal to the
thermal velocity of the cold gas and the initial $R$ was set equal
to that determined by the Coulomb repulsion measured in the
experiment at the minimum $\tau$. The calculation is shown in Fig.
\ref{fig:5} as a dotted line. The data from the experiment clearly
indicates that the recoil velocity of the pair of atoms excited in
the experiment is much less than the thermal velocity of the gas.
Note that larger collision velocities cause the recoil kinetic
energy from the Coulomb repulsion to decrease faster. This
experimental result and the fact that our calculations predict bound
states gives us confidence that we have observed bound macrodimers
\cite{Overstreet07}. The measurement for 4 different states and the
fact that the potential energy calculations predict potential wells
that support bound states makes it highly unlikely that the laser,
1.5$\,$MHz spectral bandwidth, could selectively excite pairs with
relative velocities of $< 3.4\,$m$\,$s$^{-1}$.

\section{Discussion}

The potential wells probed in our experiment depend strongly on
$\epsilon$. The resonances vanish at larger $\epsilon$ because the
potentials have become repulsive. At $\epsilon = 0
\,$mV$\,$cm$^{-1}$, our calculations do not predict any of the
prominent wells shown in Fig. \ref{fig:1} and the spectral features
are not observed. In the cases explored in this paper, the molecular
states are shifted as $\epsilon$ is increased until they interact to
form avoided crossings. As $\epsilon$ is increased further, the
states are shifted through each other so the bound states created by
the avoided crossings are destroyed. The electric fields were chosen
for our experiments so that the potentials wells were broad and deep
so that the molecules were most easily observed.

A closer look at the molecular signal in Fig. \ref{fig:3} indicates
that there may be an alignment effect. The symmetric maxima about
the peak center were observed for all the data. If there were no
alignment effect, the TOF distribution would look flat topped with a
smoothing from the thermal velocity \cite{Mons88}. We measured the
TOF distribution for the molecular resonance with the dye laser
polarization perpendicular to the TOF axis to explore this feature.
We observed no difference in the width or shape of the distribution.
We conclude that if alignment of the molecule is present, it is due
to $\epsilon$. We currently have no way of verifying this conclusion
without major modifications to the apparatus. We plan to study this
effect in the future.

To conclude, we have demonstrated the existence of bound molecular
states of Cs Rydberg atom pairs by mapping the $R$ distribution of
the atoms using the Coulomb repulsion of the ions after PFI. Our
experiments are in good agreement with Monte Carlo calculations
using no adjustable parameters of the TOF distributions and
calculations of the pair interactions of the Rydberg atoms. These
macrodimers are stabilized by $\epsilon$ and result from avoided
crossings. $\epsilon$ can be used to control the formation and
characteristics of the molecules. Macrodimers like these will be
observable in other species of cold Rydberg gases. With the higher
densities that can be achieved in an optical dipole trap, imaging
studies can be carried out to investigate the alignment effect noted
here. Now that macrodimers have been observed, their production can
be optimized and they can be trapped so these sensitive, novel
states of matter can be used for new experiments.

\section{Acknowledgements}
We acknowledge support from AFOSR (FA9550-05-0328) and ARO
(W911NF-08-1-0257).


\begin{thebibliography}{99}


\bibitem{Mourachko97} Mourachko, I.~et al. Many-body effects
in a frozen Rydberg gas. \emph{Phys.~Rev.~Lett.} \textbf{80,} 253
(1998).

\bibitem{Anderson97} Anderson, W.~R., Veale, J.~R. $\&$ Gallagher, T.~F.
Resonant dipole-dipole energy transfer in a nearly frozen Rydberg
gas. \emph{Phys.~Rev.~Lett.} \textbf{80,} 249 (1998).

\bibitem{Gould04} Tong, D.~et al. Local blockade of
Rydberg excitations in an ultracold gas. \emph{Phys. Rev. Lett.}
\textbf{93,} 63001 (2004).

\bibitem{Weidemuller04} Singer, K., Reetz-Lamour, M., Amthor, T.,
Marcassa, L.~G. $\&$ Weidem\"uller, M. Suppression of excitation and
spectral broadening induced by interactions in a cold gas of Rydberg
atoms. \emph{Phys.~Rev.~Lett.} \textbf{93,} 163001 (2004).


\bibitem{Raithel05} Cubel Liebisch, T., Reinhard, A., Berman, P.~R. $\&$
Raithel, G. Atom counting statistics in ensembles of interacting
Rydberg atoms. \emph{Phys.~Rev.~Lett.} \textbf{95,} 253002 (2005).

\bibitem{Pillet06} Vogt, T.~et al. Dipole blockade at F\"orster resonances in high
resolution laser excitation of Rydberg states of Cesium atoms.
\emph{Phys.~Rev.~Lett.} \textbf{97,} 083003 (2006).

\bibitem{Pfau07} Heidemann, R.~et al. Evidence for coherent
collective Rydberg excitation in the strong blockade regime.
\emph{Phys.~Rev.~Lett.} \textbf{99,} 163601 (2007).

\bibitem{Jaksch00} Jaksch, D.~et al. Fast quantum gates for neutral atoms.
\emph{Phys.~Rev.~Lett.} \textbf{85,} 2208 (2000).

\bibitem{Lukin01} Lukin, M.~D. et al. Dipole blockade and quantum
information processing in mesoscopic atomic ensembles. \emph{Phys.~
Rev.~Lett.~} \textbf{87,} 37901 (2001).

\bibitem{Saffmann08} Brion, E., Molmer, K. $\&$ Saffman, M. Quantum
computing with collective ensembles of multilevel systems.
\emph{Phys.~Rev.~Lett.} \textbf{99,} 260501 (2007).

\bibitem{Green00} Greene, C.~H., Dickinson, A.~S. $\&$ Sadeghpour, H.~R.
 Creation of polar and nonpolar ultra-long-range Rydberg molecules.
\emph{Phys.~Rev.~Lett.} \textbf{85,} 2458 (2000).

\bibitem{Boisseau02} Boisseau, C., Simbotin, I. $\&$ Cote, R.
Macrodimers: ultralong range Rydberg molecules.
\emph{Phys.~Rev.~Lett.} \textbf{88,} 133004 (2002).

\bibitem{Schwettmann07} Schwettmann, A., Overstreet, K.~R., Tallant, J.
$\&$ Shaffer, J.~P. Long range Cs Rydberg molecules.
\emph{J.~Mod.~Opt.} \textbf{54,} 2551 (2007).

\bibitem{Bendkowsky09} Bendkowsky, V.~et al. Observation of
ultralong range Rydberg molecules. arXiv:0809.2961.


\bibitem{Farooqi03} Farooqi, S.~M.~et al. Long-range molecular
resonances in a cold Rydberg gas. \emph{Phys.~Rev.~Lett.}
\textbf{91,} 183002 (2003).

\bibitem{Overstreet07} Overstreet, K.~R., Schwettmann, A., Tallant, J.
$\&$ Shaffer, J.~P. Photoinitiated collisions between cold Cs
Rydberg atoms. \emph{Phys.~Rev.~A} \textbf{76,} 011403(R) (2007).

\bibitem{Schwettmann06} Schwettmann, A., Crawford, J., Overstreet, K.~R.
$\&$ Shaffer, J.~P. Cold Cs Rydberg-gas interactions.
\emph{Phys.~Rev.~A} \textbf{74,} 020701(R) (2006).

\bibitem{Tallant06} Tallant, J., Overstreet, K.~R., Schwettmann, A. $\&$
Shaffer, J.~P. Sub-Doppler magneto-optical trap temperatures
measured using Rydberg tagging. \emph{Phys.~Rev.~A} \textbf{74,}
023410 (2006).

\bibitem{Mons88} Mons, M. $\&$ Dimicoli, I. Angular correlation
between photofragment velocity and angular momentum measured by
resonance enhanced multiphoton ionization detection.
\emph{J.~Chem.~Phys.} \textbf{90,} 4037 (1989).

\bibitem{Farley81} Farley, J.~W. $\&$ Wing, W.~H.,
Accurate calculation of dynamic Stark shifts and depopulation rates
of Rydberg energy levels induced by blackbody radiation. Hydrogen,
Helium, and Alkali-metal atoms. \emph{Phys.~Rev.~A} \textbf{23,}
2397 (1981).

\bibitem{He90} He, X., Li, B., Chen, A. $\&$ Zhang, C.,
 Model-potential calculation of lifetimes of Rydberg states of Alkali
atoms. \emph{J.~Phys.~B} \textbf{23,} 661 (1990).





\end{thebibliography}
\end{document}